\documentclass{article}

\usepackage{arxiv}
\usepackage[square,numbers]{natbib}
\usepackage[utf8]{inputenc} 
\usepackage[T1]{fontenc}    
\usepackage{hyperref}       
\usepackage{url}            
\usepackage{booktabs}       
\usepackage{amsfonts}       
\usepackage{nicefrac}       
\usepackage{microtype}      
\usepackage{lipsum}		
\usepackage{graphicx}
\usepackage{doi}
\usepackage{tabularx}
\usepackage{graphicx}
\usepackage{subcaption}
\usepackage{amsmath}
\usepackage{mdframed}
\usepackage{longtable}
\newcolumntype{P}[1]{>{\raggedright\arraybackslash}p{#1}}
\usepackage{comment}
\usepackage{csquotes}
\usepackage[table]{xcolor}
\usepackage{multirow}
\usepackage{array}
\usepackage{orcidlink}
\usepackage{authblk} 
\usepackage{listings}
\usepackage{soul}
\usepackage{xcolor}

\usepackage{fancyvrb}

\usepackage{pgf-pie}
\usepackage{pgfplots}
\usepackage{pgfplotstable}
\usetikzlibrary{patterns,shapes.geometric,arrows,chains,scopes,positioning, arrows.meta}
\pgfplotsset{compat=1.9}
\usepackage{seqsplit}
\usepackage{textcomp}
\usepackage{tikz}
\usetikzlibrary{trees}

\title{A Design Science Blueprint for an Orchestrated AI Assistant in Doctoral Supervision}
\author[1]{Teo Susnjak\orcidlink{0000-0001-9416-1435}}
\author[2,4]{Timothy R. McIntosh\orcidlink{0000-0003-0836-4266}}
\author[1]{Tong Liu\orcidlink{0000-0003-3047-1148}}
\author[3]{Paul Watters\orcidlink{0000-0002-1399-7175}}

\affil[1]{School of Mathematical and Computational Sciences, Massey University, Auckland, New Zealand}
\affil[2]{Cyberoo Pty Ltd, Surrey Hills, NSW, Australia}
\affil[3]{Cyberstronomy Pty Ltd, Ballarat, VIC, Australia}
\affil[4]{RMIT University, Melbourne, VIC, Australia}

\renewcommand{\shorttitle}{Orchestrated AI Assistant in Doctoral Supervision}

\begin{document}
\maketitle

\begin{abstract}
This study presents a design science blueprint for an orchestrated AI assistant and co-pilot in doctoral supervision that acts as a socio-technical mediator. Design requirements are derived from Stakeholder Theory and bounded by Academic Integrity. We consolidated recent evidence on supervision gaps and student wellbeing, then mapped issues to adjacent large language model capabilities using a transparent severity–mitigability triage. The artefact assembles existing capabilities into one accountable agentic AI workflow that proposes retrieval-augmented generation and temporal knowledge graphs, as well as mixture-of-experts routing  as a solution stack of technologies to address existing doctoral supervision pain points. Additionally, a student context store is proposed, which introduces behaviour patches that turn tacit guidance into auditable practice and student-set thresholds that trigger progress summaries, while keeping authorship and final judgement with people. We specify a student-initiated moderation loop in which assistant outputs are routed to a supervisor for review and patching, and we analyse a reconfigured stakeholder ecosystem that makes information explicit and accountable. Risks in such a system exist, and among others, include AI over-reliance and the potential for the illusion of learning, while guardrails are proposed. The contribution is an \textit{ex ante}, literature-grounded design with workflow and governance rules that institutions can implement and trial across disciplines.
\end{abstract}


\section{Introduction}
\label{sec:introduction}

The landscape of doctoral education is undergoing a significant transformation, both by the rapid diffusion of AI technologies \cite{davar2025ai} as well as an increase in the enrollment of students in online doctoral programs \cite{burrus2019predictors,melian2023online}, a trend accelerated by the recent pandemic \cite{melian2023online}. 
This shift towards virtuality has also changed the demographic profile of contemporary doctoral candidates who now tend to be working adults, often with family responsibilities, pursuing their degrees online and part-time \cite{offerman2011profile}. However, this evolution in doctoral education coincides with challenges. Historically, doctoral programs have struggled with high attrition and delayed completion rates \cite{Sowell2008Completion}, right through to the present \cite{Young2019Factors,moshtari2025supervisors}. The completion rates of doctoral degrees differ widely across countries, institutions, and disciplines, influenced by factors such as supervision quality, funding, and support services \cite{madan2021brief,spronken2018factors}. Large cross-program analyses report ten-year PhD completion rates clustering around one-half \cite{Sowell2008Completion}. The situation is exacerbated in the realm of online doctorates, where dropout rates can be up to 60--70\% \cite{rigler2017agency,marston2019issues}.

Reasons for attrition are multi-factorial and therefore complex, but recent studies have pointed to a concerning trend regarding the mental health of doctoral students in general. Instances of poor well-being, high stress, and burnout are alarmingly prevalent, increasing the risk of psychiatric disorders relative to the wider population \cite{evans2018evidence}. Newer evidence identifies proximal drivers that push students toward withdrawal: unreasonable expectations and harsh criticisms in research settings, together with anxiety and depression, where students with severe symptoms are three to five times more likely to consider leaving their programmes, combined with limited teacher training while having to take on teaching duties \cite{busch2024behind}. Systemic pressures also matter. Global survey data highlight financial precarity, harassment, and concerns about supervisory conduct as salient risks that undermine progress and retention \cite{nordling2025global}. At the institutional level, hindrance-type demands in the form of bureaucratic ambiguity, resource inadequacy, and inconsistent requirements are predictors of strain, reduced engagement, and withdrawal intentions among doctoral students \cite{acharya2024challenge}. The quality and timeliness of supervision and feedback are also repeatedly implicated with unclear or delayed feedback and power asymmetries in the supervisory relationship cited as impairing progress and well-being \cite{bearman2024enhancing,moshtari2025supervisors}. These issues have far-reaching implications, not only affecting the individuals involved but also imposing significant costs on institutions and society at large. While individuals face emotional distress and potential loss of personal and professional opportunities, institutions also suffer from the inefficient utilisation of resources together with the failure to retain talent, which ultimately affects overall advancement of knowledge and research across various fields \cite{madan2021brief,moshtari2025supervisors}.

The emergence of Large Language Models (LLMs) has already revolutionised diverse facets of human-computer interactions, from education through to research and industry in general \cite{davar2025ai}. With respect to PhD students, across 2024–2025, AI use (primarily LLM-based) is now mainstream among doctoral candidates, with on UK-based study reporting ~66\% having used AI tools, most often for proofreading (49\%), academic writing (41\%), idea generation (39\%), and literature reviews (33\%), and with 81\% of users rating the tools helpful or very helpful \cite{Akbar2025}. The same study reports its use for data analysis up to 35\% and research planning reaching 18\%  \cite{Akbar2025}. In the wider student population, the large HEPI\footnote{Higher Education Policy Institute based at Oxford, UK} 2025 survey \cite{HEPI2025} found any-AI use at 92\%, up from 66\% in 2024, reporting that 88\% have used generative AI for assessments, and that a fifth have inserted AI-generated text into assessed work. Global polling corroborates scale with 86\% of students using AI and over half doing so weekly \cite{DEC2024}, while daily use rises up to one third among postgraduate students. 

The capabilities of the new AI technologies are not only confined to academic tasks. 
In mental health, randomised trials and recent reviews show that conversational agents can deliver low-intensity support for anxiety and depression, with small to moderate short-term symptom gains relative to information-only or waitlist controls, though evidence is heterogeneous and safety reporting remains sparse \cite{hua2025scoping}. Reviews in psychiatry outline where LLMs might assist clinicians and patients with clear escalation routes \cite{stade2024large}. The proficiency of LLMs in emotion recognition and cognitive empathy is expanding, as shown by their success in ``Theory of Mind" tasks \cite{kosinski2024evaluating}, indicating also their potential to assist with addressing the increasing mental health needs of doctoral students. 

Therefore, one obvious and potential application of LLMs\footnote{We use LLMs as an umbrella term that covers generative AI systems and methods built on them. This includes agentic orchestration of multiple models, tool or function calling, retrieval, test-time compute, inference scaling, and model ensembling. References to ``LLM'' denote this broader stack, not a single model/technology.}
that remains relatively under-explored within academic research is its role as an assisting co-piloting tool for PhD supervision \cite{Iatrellis2025}, which this paper addresses through a design science approach by constructing a conceptual artefact to augment human mentorship. The traditional model of PhD supervision has primarily relied on human expertise and mentorship, but the escalating demands on academic supervisors as well as the more complex needs of doctoral students and the rapidly evolving nature of research have led to questions regarding the adequacy and efficiency of the existing human-centric supervision model \cite{bogelund2015supervisors,corner2017relationship,heath2002quantitative,madan2021brief}. 
In light of both the challenges and rapidly increasing capabilities of LLMs to mitigate them, those involved in the supervision of PhD candidates in the tertiary education sector stand to gain by acknowledging these technological developments and trends, and giving thought to how they can be optimally leveraged. The integration of these technologies into the educational framework needs to be addressed not as a distant possibility but as a present reality, thereby necessitating the design and, in a preliminary sense, an \textit{ex ante} evaluation of an artefact outlining its potential roles, benefits, and implications.

\subsection{Research Aims and Objectives}

This study adopted a design science research (DSR) approach to design and justify a conceptual blueprint for an orchestrated AI assistant and co-pilot in doctoral supervision. We did not seek to re-litigate the known shortcomings of human supervision, which are well documented \cite{heath2002quantitative,madan2021brief}. Instead, we constructed a literature-grounded artefact that harnesses model-adjacent capabilities to address recurrent gaps. Our objectives were to:
\begin{enumerate}
  \item define requirements by synthesising recent supervision pain points through the lens of Stakeholder Theory, producing a transparent severity–mitigability triage;
  \item propose a technically feasible and pedagogically sound workflow that keeps authorship and judgement with people while using orchestration of AI agent technologies, an institutional policy index, a student context store, and \emph{behaviour patches} to encode supervisory guidance at the point of use;
  \item conduct an \textit{ex ante} analysis of mitigability and risk, including bias, over-reliance, and the illusion of learning, within the governance constraints set by Academic Integrity Principles.
\end{enumerate}

Throughout, we focused on how the AI assistant functions as a socio-technical mediator and \textit{co-piloting tool}, to reconfigure information flows and responsibilities among the student, supervisor, and an institutional Graduate Research School (GRS).

\subsection{Scope and Significance of the Study}

The contribution is a \textbf{design artefact}: a blueprint comprising a conceptual framework, a process workflow, and governance rules for responsible use. Its significance is threefold.
\begin{enumerate}
  \item \textbf{Theoretical contribution.} We formalised the assistant’s role as a mediator anchored in Stakeholder Theory and bounded by Academic Integrity, offering a defensible alternative to framing an AI agent as a ``co-supervisor”.
  \item \textbf{Practical contribution.} We specified an orchestrated workflow with implementable components (retrieval grounding, expert routing, inference-time scaling, multimodality, policy index, context store) and introduced \emph{behaviour patches} that make tacit guidance explicit and auditable while keeping responsibility with people.
  \item \textbf{Guidance for future research.} We provided an \textit{ex ante} agenda for field trials and governance studies by identifying where mitigability is highest and where risks like the \textit{illusion of learning} need to be actively designed against.
\end{enumerate}


\section{Literature Review}
\label{sec:LiteratureReview}
We review the literature to first consider the nature of PhD students' identity formation, followed by a scan of typical task distributions in PhD supervisions. We then review cited inadequacies in current human-based PhD supervision arrangements at universities. This leads to an overview of the emerging capabilities of LLMs' and associated AI technologies to highlight their mitigation potential in supervision pain points, before highlighting the relationship between Bloom's Taxonomy and the identified emerging LLM capabilities, to align AI assistance with target cognitive levels.

\subsection{PhD Student's Identity Formation}
\label{subsec:IdentityFormation}
Doctoral identity is not a mindset shift but a trajectory in which candidates move from peripheral participation to accountable authorship within a community \cite{leshem2020identity,tobbell2010exploring}. This trajectory advances when students take responsibility for consequential tasks that the community recognises, such as framing a question, defending a method choice, and carrying authorship on public outputs \cite{dollarhide2013professional,tobbell2010exploring}. Identity signals cluster around routine milestones that confer recognition and agency, including confirmation reviews, first–author submissions, seminar talks, and the supervision of juniors, and these experiences, alongside all the setbacks, build belonging and resilience through practice rather than only through support \cite{acker2015struggle}. Progress is uneven across contexts because supervisory style, local lab norms, and access to networks shape who gets opportunities to perform these identity-making tasks, and international candidates often carry extra identity work due to language and unfamiliar academic codes \cite{mukminin2019acculturative,acker2015struggle}. Supervisors are thus key in this transformation journey, providing not only subject expertise, but also the validation that counters frequently present \textit{impostor syndrome} and isolation, and model the tenacity needed to navigate both the inevitable setbacks and uncertainty. This raises, therefore, central questions for our study: What is the potential role of an AI assistant in this relational process of identity formation for PhD students? Can the emerging technologies provide scaffolds that support this journey, or does their non-human nature risk exacerbating the very isolation that makes this transition so challenging?

\begin{figure}[htb!]
    \centering
    \includegraphics[width=0.75\linewidth]{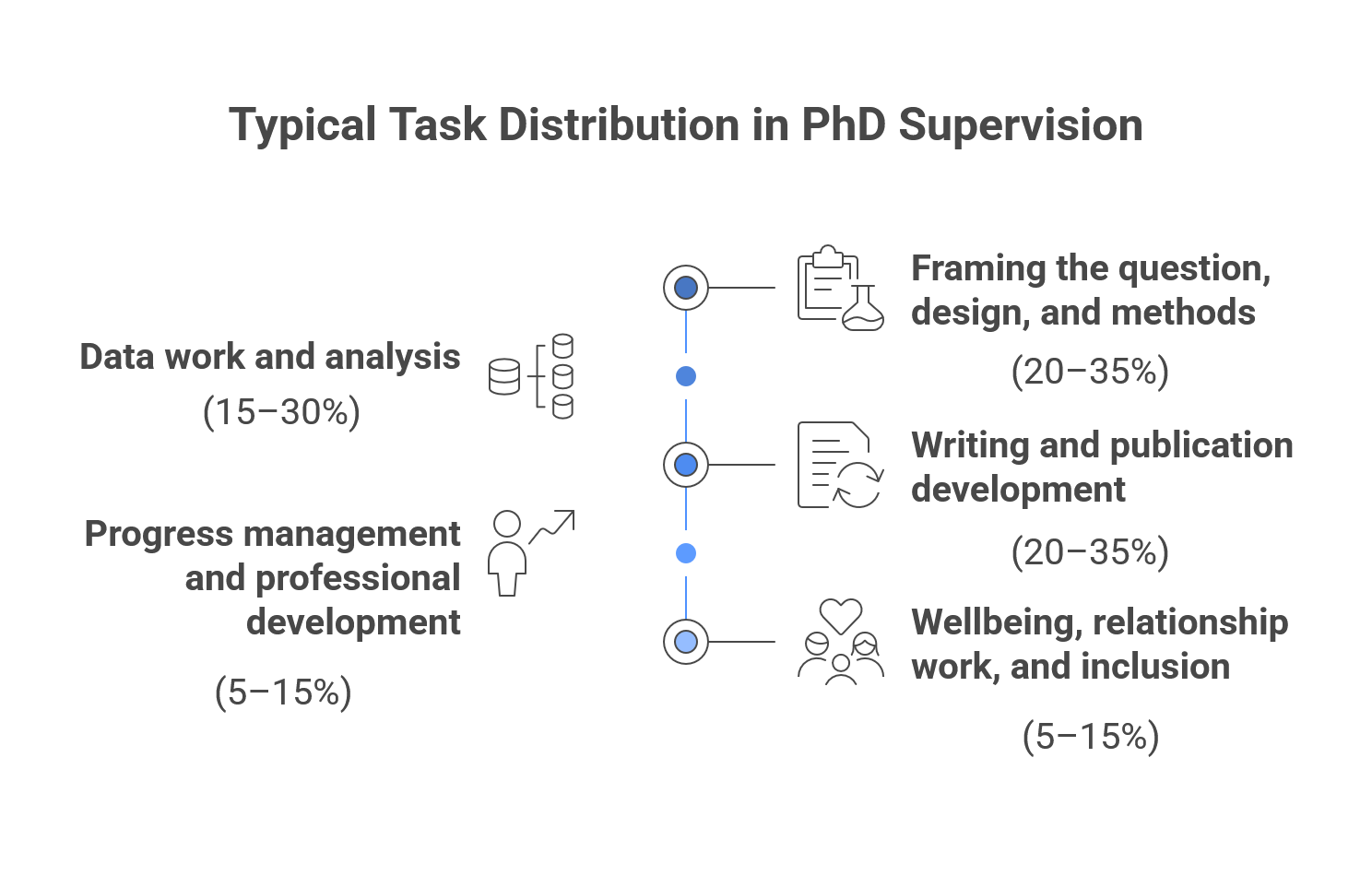}
    \caption{Estimates of typical PhD supervision tasks and effort distribution bands \cite{AitchisonLee2006,Granello2004,Bogelund2015,levecque2017work}.}
    \label{fig:effort}
\end{figure}

\subsection{Typical Task Distribution in PhD Supervision}
\label{subsec:TaskDistribution}

The distribution of supervisory work follows the same identity trajectory described above, because candidates earn recognition by taking on more of each task over time while supervisors shift from doing to coaching \cite{UKCGE2023,Bogelund2015}. There is no single split that fits every field, yet national frameworks, institutional policies, and the doctoral-writing literature point to a stable set of functions whose effort rises and falls with project stage and local norms \cite{UKCGE2023,Bogelund2015}. A consistent thread across all functions is the need to keep asking and refining the research question, since design choices, method selection, and data generation depend on the exact question being answered \cite{UKCGE2023}. We synthesise comparative guidance and workload models and present typical effort bands with brief rationales in Figure~\ref{fig:effort} and summarise them here.

\begin{itemize}
  \item \textbf{Framing the question, design, and methods (20--35\%)} defining aims, operationalising constructs, choosing methods and ethics pathways; effort peaks early and during pivots \cite{UKCGE2023,Bogelund2015}.
  \item \textbf{Data work and analysis (15--30\%)} planning data acquisition, quality checks, analysis, and interpretation across lab, field, archival, or computational pipelines \cite{UKCGE2023,Bogelund2015}.
  \item \textbf{Writing and publication development (20--35\%)} thesis chapters, papers, and responses to reviewers where iterative feedback carries the scholarship \cite{AitchisonLee2006,Granello2004}.
  \item \textbf{Progress management and professional development (5--15\%)} milestone planning, candidature reviews, networking, careers, and research integrity \cite{UKCGE2023}.
  \item \textbf{Wellbeing, relationship work, and inclusion (5--15\%)} pastoral support and timely signposting, with demand rising around setbacks and deadlines \cite{levecque2017work,UKCGE2023}.
\end{itemize}

Institutional models usually allocate a yearly envelope of hours per student (often \(\sim\)50--80) rather than prescribing within-envelope shares, which supports the banded view above and recognises stage and field dependence \cite{OxfordMPLS2024}. We therefore report bands rather than universal ratios and treat the evolving research question as the organising principle that drives movement across categories.

\subsection{Gaps and Inefficiencies in Human PhD Supervisions}
\label{subsec:FlexibilityInSupervision}

The identity and task trajectories above depend on timely recognition, clear expectations, and access to consequential work, yet recent studies show these conditions are uneven in practice. Universities have expanded supports \cite{kiley2011developments,madan2021brief,cornell2022supervisor}, but persistent gaps remain, with heavy workloads, variable communication, and inconsistent guidance linked to dissatisfaction and attrition \cite{heath2002quantitative,corner2017relationship,al2020mentorship}.
Consistent pain points have been reported recently to include slow feedback cycles and weak feedforward quality that delay progress \cite{bearman2024enhancing}; unclear or shifting expectations that heighten uncertainty and stress \cite{bearman2024enhancing,busch2024behind}; relationship strain and power asymmetries that complicate negotiation of scope, credit, and work conditions \cite{bearman2024enhancing,moshtari2025supervisors,mavrogalou2024relationship}; workload pressure and administrative demands that crowd out supervision time \cite{acharya2024challenge}; pressure to publish that distorts priorities and raises perceived risk \cite{nordling2025money}; teaching burdens with limited preparation that add to cognitive load \cite{busch2024behind}; boundary strain and emotional load on supervisors during crises and conflicts \cite{li2025phd,bearman2024enhancing}; and information overload in the literature that makes synthesis and decision-making harder \cite{arnold2023dealing,shahrzadi2024causes}. A further source of friction is policy clarity, where ambiguity about acceptable AI use and integrity expectations undermines confidence in supervision processes \cite{jin2025generative,nordling2025money}. These issues set the context for the summary of recurrent supervision problems in Section \ref{subsec:IssuesAndImpactOfGPT}. Meanwhile, these recurring inefficiencies centred around feedback, communication, and consistency are precisely the kinds of process-level problems where emerging AI technologies present targeted solutions. Early explorations of tripartite models, where an AI assistant collaborates with the student and supervisor, show promise \cite{Iatrellis2025}. However, these initial forays remain under-studied and highlight the need for a more robust, governable, and pedagogically-grounded blueprint for a comprehensive AI co-piloting system.

\subsection{LLMs and Their Emerging Capabilities}
\label{subsec:LLMCapabilities}

The previous subsection outlined recurrent gaps in human supervision. Here we survey the current LLM-adjacent capabilities that an AI assistant can draw on, without yet mapping them to specific problems. The focus is on what each technology adds in practice.
Debate continues over whether LLMs internalise a genuine ``world model" or a powerful statistical surrogate learned from vast regularities in data \cite{assran2025v}. For an agent to possess some internal representation of the world is important, as this equips an agent to understand causality and temporality, which then enables complex reasoning and planning capabilities.
For our purposes, the debate about whether AI agents inherently possess this at present is secondary. Nonetheless, effective supervision requires the assistant to operate with a usable, \textit{functional} world model of tasks, states, and constraints. This functional model does not need to exist inherently within the \textit{base} LLM's model weights; instead, it can be formed and practically emulated through the system's interaction with its surrounding components.
 Empirical results show that, under light orchestration, language models can maintain a goal across steps and choose actions that advance a task \cite{yao2023react,jin2023cladder}. This supports treating the system as one where the base LLM supplies compressed abstractions over language, sufficient for reasoning and following instructions, while the surrounding components, such as memory, tools, and grounded data, supply the time-aware state and verification that constitute a \textit{practical} world model \cite{feng2025embodiedaillmsworld}.
Table~\ref{tab:LLMCapabilities} summarises these key capabilities. Each one addresses a known limitation of a standalone LLM.

\begin{table}[htb!]
\centering
\caption{Key LLM-adjacent and augmentation technologies, and what each adds in practice}
\label{tab:LLMCapabilities}
\renewcommand{\arraystretch}{1.25}
\begin{tabular}{p{0.30\linewidth} p{0.62\linewidth}}
\toprule
\textbf{LLM-augmentation Capability} & \textbf{What it adds} \\
\midrule
Retrieval-Augmented Generation (RAG) & Grounds answers in a defined corpus with citations and source backlinks. Local policy awareness and auditable claims \cite{lewis2020retrieval}. \\
\midrule
Cache-Augmented Generation (CAG) & Supports stable long conversations by caching salient turns and snippets, and thus retaining context and conversational coherence over time \cite{chan2025don}. \\
\midrule
Temporal Knowledge Graphs (TKGs) & 
Provide a time-aware state and reasoning over a multi-year candidature: entities–relations–timestamps for provenance, sequencing, and constraint checks. Enables temporal retrieval (``what changed, when''), early warning for slippage, and simple trajectory forecasting. \cite{cai2023temporal} \\
\midrule
Tool use and Agentic AI & Access to search, code and its execution, validators, and calendars. Visible intermediate steps and reproducible actions via logs \cite{schick2023toolformer}. \\
\midrule
Inference-time scaling & Extra reasoning on hard prompts with lightweight self-checks. Compute is tied to task difficulty for higher-quality responses \cite{snell2025scaling}. \\
\midrule
Mixture-of-Experts (MoE) & Specialised routing for subject-domains, specific LLM capabilities (e.g. wellbeing and coaching) or input types. Higher effective capacity without linear cost growth \cite{shazeer2017outrageously}. \\
\midrule
Ensembles and self-consistency & Ability to moderate and reconcile conflicting and subjective responses from different LLMs, and voting to stabilise outputs with high variability. \\
\midrule
Multimodality (text, tables, figures, audio, video) & Direct reading and processing of non-textual inputs. \\
\midrule
Structured memory and logging & Persistent state for decisions, rationales, and versions. Replay and audit of who changed what and why. \\
\bottomrule
\end{tabular}
\end{table}

\paragraph{How these parts fit together.} The stack starts with an agentic orchestration layer that acts as a goal-directed router. It selects the right model for a task, chooses which tools to call, sequences the steps, and keeps an audit trail. The orchestrator grounds answers using RAG as needed, not only on the student’s subject corpus but also on local artefacts like ethics approvals and supervisor notes. CAG keeps long dialogues stable, which otherwise lose coherence. Because a PhD unfolds over years, TKGs provide the time-aware state: who did what, when milestones changed, and which dependencies must hold. This enables ``what changed/what’s next" queries and early warnings for slippage. Over this shared state, the system executes tool use, routes niche questions via MoE, and applies inference-time scaling or ensembles when a problem is hard, subjective or contested. The result is a powerful, context-aware assistant for co-piloted supervision, empowered by a cluster of enabling technologies, with the potential to not only reduce routine work but also enhance supervision.

\subsection{Mapping the Orchestrated System to Bloom's Taxonomy}
\label{subsec:BloomsTaxonomy}

The previous subsection surveyed the capabilities of the orchestrated assistant at a technical level. Here, we align those capabilities to Bloom’s cognitive levels to demonstrate how the system can route support based on the kind of thinking a task is meant to build \cite{Bloom1956,krathwohl2002revision}.

\begin{enumerate}
  \item \textbf{Remember:} External tooling like \textbf{RAG} transforms this level from simple recall of LLM's pre-trained knowledge to the auditable retrieval of verifiable facts from a defined corpus, such as institutional policies or a target literature index \cite{lewis2020retrieval}.

  \item \textbf{Understand:} The system supports understanding by generating summaries and explanations that are explicitly grounded in the sources provided by RAG. \textbf{CAG} and structured memory ensure this understanding remains consistent across long, multi-session conversations.

  \item \textbf{Apply:} \textbf{Agentic tool use} elevates this from suggestion to execution across \textbf{multimodal} data. The assistant can be tasked to apply a specific method by generating and running code, with the intermediate steps and final outputs remaining visible and auditable for student inspection \cite{schick2023toolformer}.

  \item \textbf{Analyse:} 
  Contrastive reads over retrieved evidence support comparison and error framing. \textbf{Ensembles} surface alternative interpretations before a line of argument is chosen; \textbf{inference-time scaling} allocates extra reasoning to probe disagreements and expose hidden assumptions. Where order and timing matter, \textbf{TKGs} enable temporal analyses—diffs across versions, event-sequence checks, and trend detection over the candidature timeline. \cite{cai2023temporal}

 \item \textbf{Evaluate:} 
  Rubric-guided critique uses retrieved standards and exemplars. Assistant behaviour can be adjusted to a question-led mode that maintains a Socratic dialogue.  \textbf{MoE} routes domain questions to specialist components; \textbf{inference-time scaling} audits borderline cases. \textbf{TKGs} add provenance- and policy-aware conformance checks against an explicit, auditable history \cite{shazeer2017outrageously,cai2023temporal}.

  \item \textbf{Create:} The assistant acts as a co-creation partner, using its generative capabilities to propose options, draft text, and offer counterexamples. However, the final authorial and accountable act of synthesis remains with the candidate, in line with current academic integrity guidance \cite{flanagin2023nonhuman}.
\end{enumerate}


\section{Research Methodology: A Design Science Approach}
\label{sec:Methodology}

This study proposes a blueprint for integrating LLMs as co-pilot tools in doctoral supervision \cite{vaishnavi2004design} using the DSR approach. The contribution is a design artefact comprising a conceptual framework and process model that augments human-centred supervision while preserving human oversight. We followed DSR cycles of rigour (literature and theory synthesis) and design (iterative refinement against stated constraints). We evaluated the artefact \textit{ex ante} using five explicit criteria grounded in prior research: problem relevance, theoretical alignment, internal coherence, governance feasibility, and anticipated utility. No new empirical data were collected for this study.

\subsection{Methods}
\label{subsec:ResearchDesign}

\paragraph{Phase 1: Literature grounding.} We conducted a targeted review (Section~\ref{sec:LiteratureReview}) to describe the formation of doctoral identity and task distribution, summarised recent supervision gaps, and scoped the emerging capabilities of LLM/AI technologies. This review provided an explicit, literature-grounded baseline for the artefact’s design.

\paragraph{Phase 2: Artefact design and justification.} We anchored our design choices in two theoretical lenses. First, Stakeholder Theory treated the PhD candidate, the supervisor, and the Graduate Research School (GRS) as parties with legitimate claims whose interests should be balanced \cite{freeman2010strategic,donaldson1995stakeholder}. The AI assistant was not treated as a stakeholder in the normative sense because it has no legitimate claims or moral standing \cite{phillips2003stakeholder,mitchell1997toward}; we treated it as a socio-technical mediator that reshapes interactions under explicit human accountability \cite{orlikowski200810,fassin2009stakeholder,ICMJE2024}. Second, Academic Integrity Principles provided the non-negotiable governance boundaries for acceptable use \cite{sotiriadou2020role}.

\paragraph{Ex ante criteria application.} We applied our stated criteria as follows. \textit{Problem relevance:} we consolidated recent, high-severity supervision pain points in Section~\ref{subsec:FlexibilityInSupervision}. \textit{Theoretical alignment:} the artefact aligned with the identity–tasks account and Bloom-level routing (Sections~\ref{subsec:IdentityFormation} and \ref{subsec:BloomsTaxonomy}). \textit{Internal coherence:} roles and process flows were fixed to the three human stakeholders with the AI assistant framed as a socio-technical mediator. \textit{Governance feasibility:} the design’s approach to authorship, disclosure, and provenance followed current integrity guidance. \textit{Anticipated utility:} Table~\ref{tab:LLMImpactRevised} functions as an issue catalogue indicating where orchestration would plausibly reduce latency and inconsistency while preserving human authorship. With respect to scope, the evaluation was document-based and conceptual. We did not present user studies, field trials, or quantitative benchmarking. Outcome validation is outlined as future work; the current contribution is a literature-grounded artefact and its ex-ante assessment against the criteria above.


\section{Analysis: Stakeholders and Pain Points}
\label{sec:Analysis}

The methodology located our artefact within Stakeholder Theory and Academic Integrity Principles. We now make that stance operational. First, we specify who the primary stakeholders are, what legitimate claims and duties they hold, and how their relations are structured. We then read recurrent pain points as breakdowns in those relations.

\subsection{The Traditional Stakeholder Triad}
\label{subsec:StakeholderTheory}

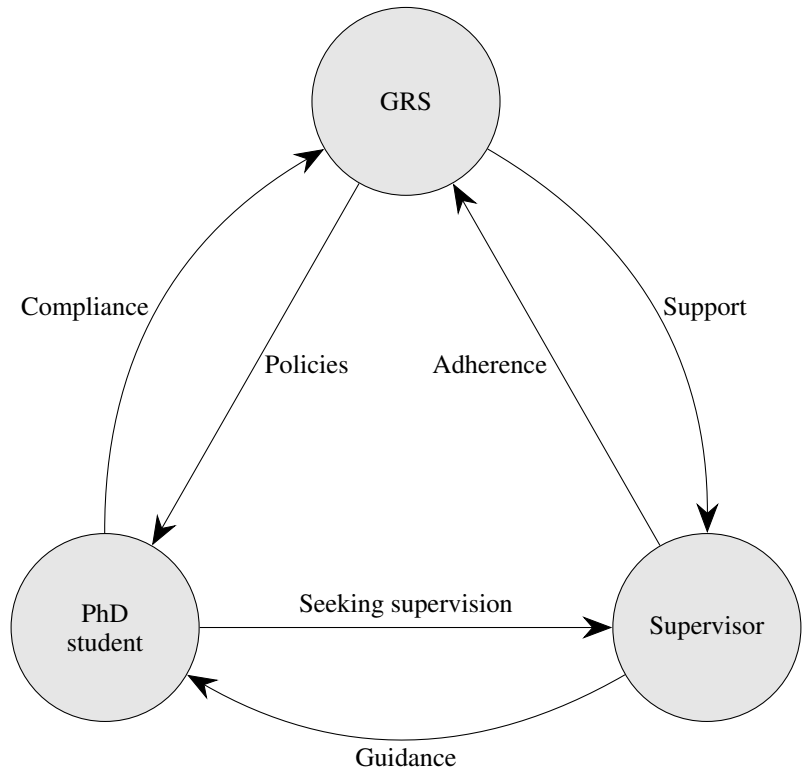
\begin{figure}[!t]
	\centering
	\begin{tikzpicture}[
		vertex/.style={circle, draw, fill=gray!20, minimum size=2.5cm, align=center},
		>={Stealth[length=4mm]}
		]
		\node[vertex] (phd) at (0,0) {PhD\\student};
		\node[vertex] (supervisor) at (8,0) {Supervisor};
		\node[vertex] (grs) at (4,7) {GRS};
		
		\draw[->] (phd) -- node[midway, above] {Seeking supervision} (supervisor);
		\draw[->] (supervisor) edge[bend left] node[midway, below] {Guidance} (phd);
		
		\draw[->] (grs) -- node[midway, right] {Policies} (phd);
		\draw[->] (phd) edge[bend left] node[midway, left] {Compliance} (grs);
		
		\draw[->] (grs) edge[bend left] node[midway, right] {Support} (supervisor);
		\draw[->] (supervisor) -- node[midway, left] {Adherence} (grs);
	\end{tikzpicture}
	\caption{Stakeholders and their relationships in the context of PhD supervision}
	\label{fig:PhDSupervisionStakeholders}
\end{figure}

Guided by Stakeholder Theory \cite{freeman2010strategic,donaldson1995stakeholder}, we treat three primary stakeholders as holding legitimate claims over doctoral supervision: the PhD candidate, the supervisor, and the GRS. Their interactions form three relation classes that organise routine supervision work (Fig.~\ref{fig:PhDSupervisionStakeholders}).

\begin{enumerate}
\item \textbf{Pedagogic relation (candidate $\leftrightarrow$ supervisor).} The candidate seeks supervision and submits work and queries for guidance. The supervisor provides formative feedback, models practice, and assures quality under competing role demands \cite{carr2010supervision,hemer2012informality}.
\item \textbf{Policy–compliance relation (candidate $\leftrightarrow$ GRS).} The GRS sets candidature requirements, integrity, and ethics policy. The candidate complies, reports progress, and receives procedural support \cite{kiley2011developments}.
\item \textbf{Capacity–assurance relation (supervisor $\leftrightarrow$ GRS).} The GRS provides training, workload rules, and institutional support. The supervisor adheres to policy and quality assurance.
\end{enumerate}

This structure gives each party clear claims and duties, and it shapes salience in practice, since deadlines raise urgency and formal authority shapes perceived power within the triad. When these relations fail, misalignment appears as delays, expectation drift, or role strain. The next subsection catalogues those failures as pain points. Secondary parties such as funders and external examiners form the broader context, while important, are not modelled here \cite{malfroy2011impact,mogaji2021stakeholders}.

\subsection{Mapping Pain Points to the AI Assistant's Capabilities}
\label{subsec:IssuesAndImpactOfGPT}

Having defined the stakeholder relations and framed pain points as failures of their quality, we now map these recurrent issues to the specific affordances of the orchestrated AI assistant. The analysis in Table~\ref{tab:LLMImpactRevised}, feeds the study’s \textit{ex ante} evaluation by estimating mitigability, indicating severity, and flagging intervention risks. In interpreting the table, the mitigation potential is rated on a four-point scale from \emph{None} (0) to \emph{High} (3). We assign a \emph{severity} score \(S\in\{1,2,3\}\) using \(S=\mathrm{round}\big((P+C)/2\big)\), where \(P\) is prevalence and \(C\) is consequence.
\noindent\textit{Risk column.} Risk reflects the likelihood and impact of \emph{new} harms introduced by the assistant, distinct from the severity \(S\) of the underlying issue. We score Low/Medium/High using: privacy and surveillance exposure; shifts in agency or power; integrity and policy drift; gaming and metric fixation; bias and disparate impact. Low = limited new exposure with strong guardrails; Medium = meaningful exposure that is mitigable in routine use; High = credible exposure in sensitive domains even with guardrails. Full workings of the above and justifications from literature can be found in the Supplementary Materials to keep the triage, claims and risk ratings transparent.

\begin{table}[hbtp]
\centering
\caption{Recurrent doctoral support and supervision issues, mapped to AI co-pilot affordances, with mitigation, severity, and risk across stakeholders}
\label{tab:LLMImpactRevised}
\small
\renewcommand{\arraystretch}{1.2}
\begin{tabular}{p{0.26\linewidth}p{0.02\linewidth}p{0.12\linewidth}p{0.36\linewidth}p{0.08\linewidth}p{0.05\linewidth}}
\hline
\textbf{Issue} & \textbf{S} & \textbf{Stakeholder(s)} & \textbf{Co-pilot capability mapping} & \textbf{Mitigation} & \textbf{Risk} \\
\hline
Mental-health strain linked to research/teaching pressures and progress reporting \cite{busch2024behind,levecque2017work} & 3 & Candidate & Progress scaffolds, gentle pacing prompts, weekly check-ins that surface stuck points for supervisor discussion and follow-up & 1 & H \\
\hline
Harsh criticism and unreasonable expectations \cite{busch2024behind} & 3 & Candidate; Supervisor & Tone rewriter for feedback, constructive-framing templates, expectation-calibration checklists tied to agreed milestones & 2 & M \\
\hline
Pressure to publish and ``publish-or-perish” expectations \cite{nordling2025money} & 3 & Candidate; Supervisor & Journal-fit scoping/recommendations; paper reviews/proofreading; submission-readiness checklists (ethics, data, reporting); reviewer-response drafting; pipeline planners; & 2 & H \\
\hline
Teaching burden with limited training \cite{busch2024behind} & 2 & Candidate & Lesson-prep aids, rubric templates, micro-teaching scripts, common Q\&A generation with references & 2 & L \\
\hline
Emotional load, boundary strain, and reputational risk \cite{li2025phd,bearman2024enhancing} & 2 & Supervisor & Meeting-note summaries, expectation-setting templates, boundary-setting scripts, escalation decision aids, reflective prompts & 1 & M \\
\hline
Feedback timeliness and feedforward quality \cite{bearman2024enhancing} & 3 & Candidate; Supervisor & Draft-level formative comments aligned to rubrics; actionable next-step plans; meeting-minute extraction to surface agreed actions; pre-submission self-checks that flag missing coverage against criteria & 3 & M \\
\hline
Power asymmetry in feedback and working conditions \cite{bearman2024enhancing,moshtari2025supervisors} & 3 & Candidate; Supervisor; GRS & Policy-aligned feedback prompts, transparent contribution logs, escalation/reporting signposting; human oversight remains central & 1 & H \\
\hline
Challenge–hindrance demands and bureaucratic load \cite{acharya2024challenge} & 2 & Candidate; Supervisor & Checklist automation, deadline trackers, form-filling drafts, milestone planners  & 2 & M \\
\hline
Supervisory style inconsistency and relationship strain \cite{mavrogalou2024relationship} & 2 & Candidate; Supervisor & Supervision compacts, agenda builders, negotiation scripts, minutes with action extraction and follow-ups & 2 & L \\
\hline
Synthesis and academic writing load (esp. EAL) \cite{khalifa2024using} & 2 & Candidate & Outline/argument mapping; lit-synthesis tables; cohesion/register control; citation/refs checking; journal- and rubric-aligned revision plans & 3 & M \\
\hline
Funding precarity and financial stress \cite{nordling2025money} & 3 & Candidate; GRS & Scholarship search scaffolds and application drafting, budget templates, grant boilerplate drafting; (does not change availability) & 1 & L \\
\hline
Harassment and hostile climates \cite{nordling2025money} & 3 & Candidate; Supervisor; GRS & Policy checklists, confidential reporting workflow signposting, compliance reminders; enforcement remains human & 1 & H \\
\hline
Information overload in the literature \cite{arnold2023dealing,shahrzadi2024causes} & 2 & Candidate & Retrieval-augmented clustering, contrastive summaries, synthesis tables with source backlinks, ability to ``talk across'' multiple documents & 2 & M \\
\hline
Ambiguity about acceptable AI use and integrity \cite{jin2025generative,nordling2025money} & 2 & GRS; Supervisor; Candidate & Exemplars of allowed use, disclosure prompts, policy-aware checklists, AI-literacy materials for staff/students & 2 & M \\
\hline
Supervision capability and milestone quality support \cite{bearman2024enhancing} & 1 & Student; Supervisor; GRS & Milestone rubric packs and exemplar banks; student-set goal and task completion assessments for progress summaries; supervisor CPD micro-modules; panel calibration notes; prompts in progress-report templates that nudge feedforward actions; no access to private supervisor feedback & \textbf{3} & \textbf{H} \\
\hline
\end{tabular}
\begin{flushleft}\footnotesize
\textbf{S} (Severity): \(S=\mathrm{round}\!\big((P+C)/2\big)\), where \(P\) maps survey prevalence to \(\{1,2,3\}\) and \(C\) maps documented consequences (e.g., links to intent-to-leave, sustained wellbeing detriment, cross-stakeholder harm) to \(\{1,2,3\}\). Examples informing \(P\) bands in recent global data include: financial pressure (\(\approx42\%\)), pressure to publish (\(\approx41\%\)), and harassment/discrimination (\(\approx43\%\)) \cite{nordling2025money}. Example informing \(C\): severe anxiety/depression associated with a \(\sim3\text{--}5\times\) increase in intent-to-leave \cite{busch2024behind}. Evidence on information-overload prevalence and impacts is drawn from scholarly contexts \cite{arnold2023dealing,shahrzadi2024causes}. \emph{Note:} for ``ambiguity about acceptable AI use,” a headline statistic that \(\sim64\%\) of students want more guidance \cite{nordling2025money} does not automatically set \(P=3\); we scope \(P\) to cases where ambiguity materially impairs decisions or creates integrity risk (see Supplementary Methods for banding rules).
\end{flushleft}
\end{table}

\paragraph{Interpretation:}
The analysis in Table~\ref{tab:LLMImpactRevised} supports a triage for deploying an AI assistant in doctoral education. The most severe issues (\(S=3\)) comprising mental-health strain, harassment, and systemic power asymmetries, are human and structural. AI mitigation is limited but not trivial; the appropriate role is early detection, documentation, and confidential routing to human support. This yields a sequencing principle for institutional strategy. First, use the orchestrated assistant’s agentic capabilities and MoE routing to call institutionally approved domain-specific models for pastoral-care screening and triage, which detect basic emotion and risk cues and then signpost or escalate, to automate and scaffold high-mitigation, moderate-risk tasks that consume staff capacity, which reduces routine feedback and administrative load. Then redirect the recovered time and attention to the high-severity relational challenges that require human judgement and policy \cite{bearman2024enhancing,busch2024behind}.

The assistant is most effective on issues with high mitigability, such as feedback timeliness \((S=3,\ \text{mitigation}=3)\), milestone and progress support \(3\), information overload \((S=2)\), and writing load \((S=2)\).
The assistant is most effective on issues with high mitigability, such as feedback timeliness (\(S=3\), mitigation=3), information overload (\(S=2\)), and writing load (\(S=2\)). The technical architecture matters. RAG provides verifiable, cited summaries from curated sources and reduces hallucination risk; ensembles and self-consistency stabilise outputs and can pit outputs of different LLMs against each other for moderation; inference-time scaling allocates more computation to hard problems; MoE routing directs methodological queries to specialist domain-specialist components; multimodality supports reasoning over figures, tables, and code, as well as audio and video media. This \textit{always-on} availability becomes a potential lifeline for online and part-time students who face feedback latency. The GRS’s role is enabling rather than policing by providing rubric packs, exemplar banks, and clear AI-use guidance while keeping accountability with supervisors and candidates. 

Risk scores in the table denote new exposure created by a potential assistant. High-risk categories persist where surveillance creep, policy drift, or gaming are plausible (e.g., mental-health workflows, harassment reporting, publish-or-perish dynamics, and threshold-based supervision if misused).
Guardrails—consent defaults, role-based access, audited queries, and human veto reduce, but do not remove these hazards.
The potential net effect is smoother processes with guarded automation so scarce human attention is spent where it changes outcomes.


\section{Artefact: The Orchestrated Assistant for Supervision}
\label{sec:TechnicalFeasibility}

Having triaged supervision issues and identified where an assistant has high mitigability, we now instantiate the design as a concrete workflow. Again, the artefact is not a single LLM agent but an \emph{orchestrated assistant} that binds a base LLM with other specialist models as well as to tools, retrieval, routing mechanisms, and memory, under human oversight.

\subsection{An Orchestrated Workflow}
\label{subsec:IntegratingGPT}

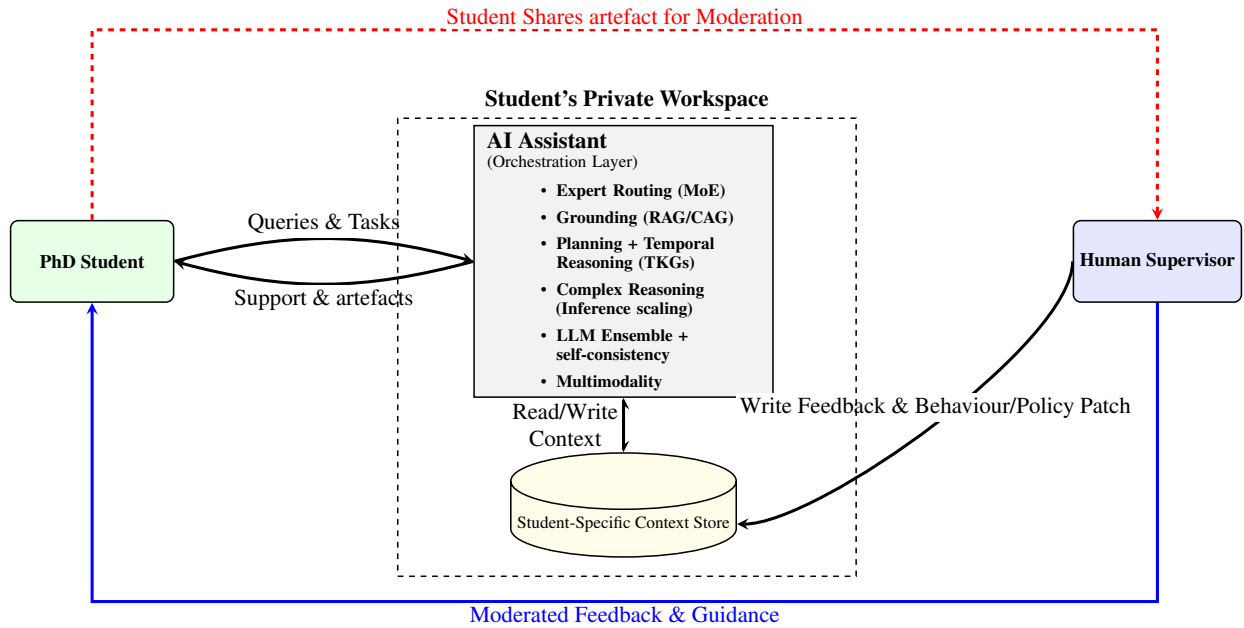
\begin{figure}[htb]
    \centering
    \resizebox{\columnwidth}{!}{
        \begin{tikzpicture}[
            node distance=2cm and 5.5cm, 
            entity/.style={rectangle, rounded corners, draw, thick, minimum height=1.5cm, minimum width=3cm, text centered, font=\bfseries},
            human/.style={entity, fill=blue!10},
            student/.style={entity, fill=green!10},
            ai_assistant/.style={rectangle, draw, thick, fill=gray!10, minimum height=5cm, minimum width=5.5cm, align=left, text width=5cm, font=\small},
            database/.style={cylinder, draw, thick, shape border rotate=90, aspect=0.25, fill=yellow!10, minimum height=1.5cm, minimum width=2cm, text centered, font=\small},
            workspace/.style={rectangle, draw, dashed, thick,  inner sep=120pt}, 
            arrow/.style={ultra thick, ->, >=stealth, font=\small\bfseries},
            label_text/.style={fill=white, align=center, inner sep=2pt, font=\large}
        ]

        \node[student] (student) {PhD Student};
        \node[ai_assistant, right=of student] (ai) {
            \textbf{\large AI Assistant} \\ (Orchestration Layer) \\[0.25em]
            \begin{itemize}
                \item \textbf{Expert Routing (MoE)}
                \item \textbf{Grounding (RAG/CAG)}
                \item \textbf{Planning + Temporal Reasoning (TKGs) }
                \item \textbf{Complex Reasoning (Inference scaling)}
                \item \textbf{LLM Ensemble + self-consistency}                
                \item \textbf{Multimodality}
            \end{itemize}
        };
        \node[database, below=1cm of ai] (db) {Student-Specific Context Store};
        \node[human, right=of ai] (supervisor) {Human Supervisor};

        \draw[arrow] (student.east) to[bend left=15] node[midway, above, label_text] {Queries \& Tasks} (ai.west);
        \draw[arrow] (ai.west) to[bend left=15] node[midway, below, label_text] {Support \& artefacts} (student.east);

        \draw[arrow, <->] (ai.south) -- node[midway, left, label_text, align=center] {Read/Write\\Context} (db.north);

        \draw[arrow, red, dashed] (student.north) |- ++(0,3.5) -| node[pos=0.25, above, label_text] {Student Shares artefact for Moderation} (supervisor.north);

        \draw[arrow] (supervisor.west) .. controls +(south:1.5) and +(east:1.5) .. node[midway, above, label_text] {Write Feedback \& Behaviour/Policy Patch} (db.east);

        \draw[arrow, blue] (supervisor.south) |- ++(0,-5.5) -| node[pos=0.25, below, label_text] {Moderated Feedback \& Guidance} (student.south);

        \begin{scope}
            \node[workspace, 
                xshift=280pt,   
                yshift=-45pt  
                ] (ws_box) {};
        \end{scope}
        \node[above, font=\bfseries\large] at (ws_box.north) {Student's Private Workspace};

        \end{tikzpicture}
    }
\caption{A proposed workflow segment for AI-assisted PhD supervision. The diagram depicts the student–AI loop and the student-initiated moderation loop with the human supervisor, where supervisor patches persist and steer later responses.
Students can set goals and thresholds that drive optional progress summaries.}

    \label{fig:ModernAIWorkflow}
\end{figure}

\paragraph{Model boundary.} Figure~\ref{fig:ModernAIWorkflow} shows a high-frequency segment of the orchestration.
It covers the private student–assistant loop and a student-initiated moderation loop focused on a specific artefact (for example, a literature subsection, a code snippet, an interpreted figure, or a brainstorming output).
This loop is mapped in Section~\ref{subsec:BloomsTaxonomy}, and supports tasks from remembering through creating by combining multiple LLM-adjacent technologies.
Students can also set task goals and thresholds in the private loop and can revoke or change them at any time.
When a goal crosses a specified completion threshold, the assistant detects this and prepares a student-curated summary for notifying the supervisor.
We model only this interaction segment because it carries most day-to-day supervision friction and it demonstrates orchestration, provenance, and behaviour patches.
Supervisor-initiated queries and GRS-level flows are part of the blueprint but are out of scope to keep the feedback path clear.
The approach generalises to those variants using the same primitives of routing, RAG and TKG state, tool logs, and human veto.

\paragraph{Assistant capabilities in this segment.} The assistant is an LLM-driven orchestration layer. It performs expert routing via MoE, grounds answers with RAG over the thesis corpus, a curated literature index, and a GRS policy index; maintains conversational coherence via CAG and time-aware reasoning using TKGs; invokes multimodal tools and applies inference-time checks (self-consistency / small ensemble) on harder prompts. The assistant is able to reason and assess completion levels of work as configured by the student. All actions and sources are logged.

\paragraph{Student-specific context store.} A private context store (comprising multiple databases to enable RAG/TKGs) maintains profiles, research descriptions, documents, rolling summaries, accepted decisions, and a ledger of artefacts with citations and verification outcomes. Timestamps are logged to enable the TKGs to reason and assist with planning using the temporal domain. Students can view, edit, or purge items. This supports continuity so responses are state-aware, tracking question maturity, design readiness, data readiness, ethics risk, and citation coverage.

\paragraph{Private loop.} Within the private loop the AI assistant selects the simplest capable route. Discipline queries use expert routing with RAG over the relevant literature and thesis materials. Policy queries use RAG over the GRS-managed index. Multimodal tasks use specialist LLMs that parse non-textual inputs. Narrow wellbeing and coaching prompts invoke an institution-approved screening model that offers reflection and signposting only; it does not deliver counselling or diagnosis and follows consent and escalation rules.

\paragraph{Student-initiated moderation loop.} Moderation is opt-in per response artefact. The student triggers (or sets completion thresholds) the forwarding of an assistant output to the supervisor. The supervisor returns feedback and/or corrections, and may attach a \emph{behaviour patch}, which is a concise constraint that steers future answers (e.g., required sources, preferred methods, scope limits, phrasing tone, or exclusions). A patch can also set a questioning mode so the AI assistant asks the student more frequently before it answers. The system records both the feedback and the patch as an \emph{assistant policy update} in the context store. Subsequent answers apply the update and cite the relevant sources. The moderated artefact is updated, and the student private-loop continues. The supervisor remains the final arbiter of AI-generated content and of assistant behaviour at programme gates.

\subsection{Reconfigured Stakeholder Ecosystem}
\label{subsec:ReconfiguredEcosystem}

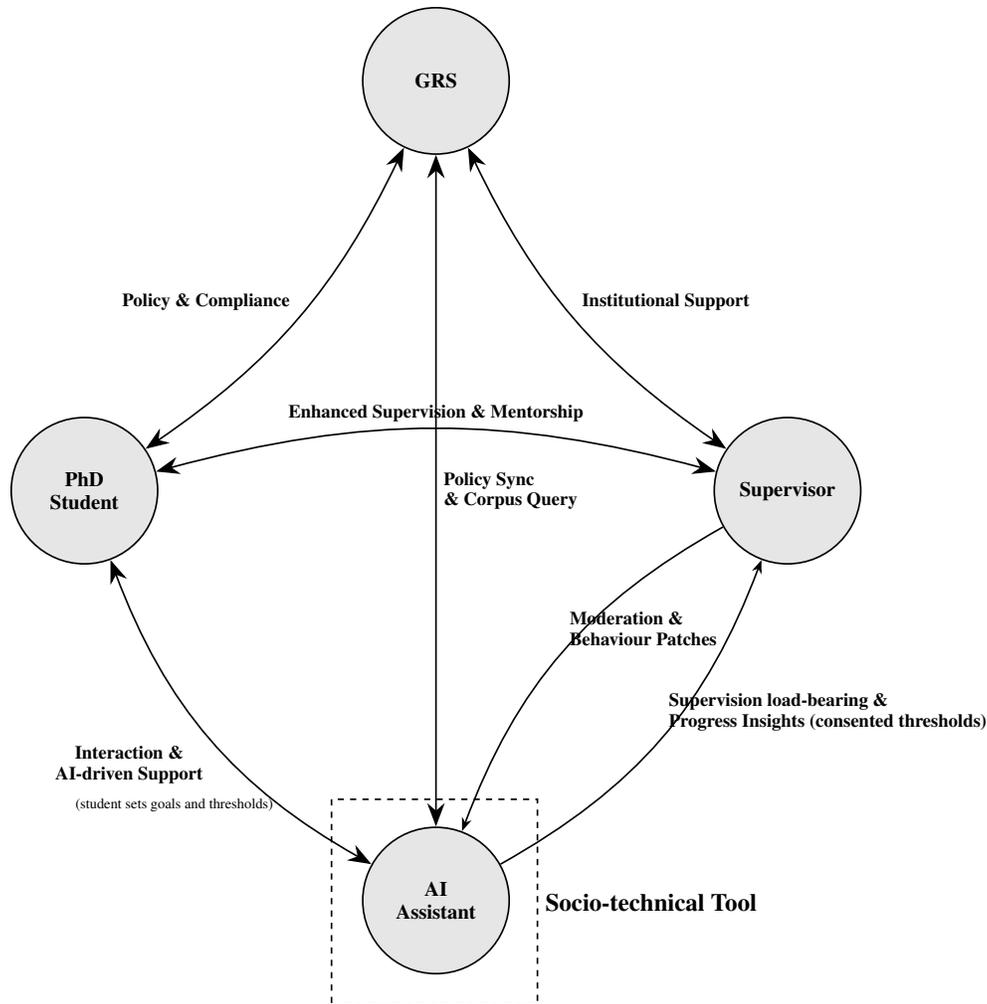
\begin{figure}[htb]
	\centering
	\resizebox{0.8\columnwidth}{!}{
		\begin{tikzpicture}[
            workspace/.style={rectangle, draw, dashed, thick,  inner sep=50pt}, 
			vertex/.style={circle, draw, thick, fill=gray!20, minimum size=2.5cm, align=center, font=\bfseries},
			arrow/.style={-Stealth, thick, font=\small\bfseries},
			>={Stealth[length=4mm]}
			]
			\node[vertex] (phd) at (0,0) {PhD\\Student};
			\node[vertex] (supervisor) at (12,0) {Supervisor};
			\node[vertex] (grs) at (6,7) {GRS};
			\node[vertex] (ai) at (6,-7) {AI\\Assistant}; 
			
			\draw[arrow, <->] (phd) to[bend left=15] node[midway, above] {Enhanced Supervision \& Mentorship} (supervisor);
			\draw[arrow, <->] (grs) to[bend left=15] node[midway, above left] {Policy \& Compliance} (phd);
			\draw[arrow, <->] (grs) to[bend right=15] node[midway, above right] {Institutional Support} (supervisor);

			\draw[arrow, <->] (phd) to[bend right=20] node[midway, below left, align=center] {Interaction \&\\AI-driven Support}
              node[pos=0.75, left, font=\scriptsize] {(student sets goals and thresholds)}
                (ai);
			
			\draw[arrow, <->] (grs) -- node[midway, right, align=left] {Policy Sync\\\& Corpus Query} (ai);
			\draw[arrow] (supervisor) to[bend right=20] node[midway, above right, align=left] {Moderation \&\\Behaviour Patches} (ai);
            \draw[arrow] (ai) to[bend right=20] node[midway, above right, align=left] {Supervision load-bearing  \& \\Progress Insights (consented thresholds)} (supervisor);
            
            \begin{scope}
                \node[workspace, 
                    xshift=170pt,   
                    yshift=-200pt  
                    ] (ws_box) {};
            \end{scope}
            \node[right, font=\bfseries\large] at (ws_box.east) {Socio-technical Tool};

		\end{tikzpicture}
	}
	\caption{Reconfigured stakeholder ecosystem with the AI assistant acting as a socio-technical mediator.}
	\label{fig:PhDSupervisionStakeholdersGPT}
\end{figure}

The introduction of the AI assistant reconfigures the relationships  (Figure~\ref{fig:PhDSupervisionStakeholdersGPT} with edges consolidated compared to the initial version, to keep the figure legible) between the three stakeholders \cite{phillips2003stakeholder,orlikowski200810,fassin2009stakeholder}.
The assistant changes how information moves and when work happens meaning that
tacit detail becomes explicit in patches and logs.
Delays shorten because rules and feedback appear at the point of use, while the roles of student, supervisor, and GRS remain the same. Instead, the relations gain clarity, timeliness, and audit trails.

\paragraph{Pedagogic relation (candidate $\leftrightarrow$ supervisor)}
    Routine scaffolding shifts to the assistant, allowing meetings to focus on judgement and identity work. Supervisor guidance is also encoded in assistant behaviour via patches that adjust the assistant's subsequent responses. This makes supervisor expectations visible, and potentially reduces delays and lowers the risk of harsh feedback delivered late.

\paragraph{Policy–compliance relation (candidate $\leftrightarrow$ GRS)}
A GRS policy index answers queries for students and also supports GRS staff who write and maintain policy.
The assistant can query across the policy corpus with RAG to surface relevant clauses, find conflicts, and trace changes.
Ambiguity reduces because rules can be applied and checked for compliance automatically, and policy authors can validate coverage and consistency more efficiently.
Student context is not exposed during GRS queries and only the policy corpus is searched.
This reduces clarity gaps identified in literature and reduces avoidable bureaucracy.

\paragraph{Capacity–assurance relation (supervisor $\leftrightarrow$ GRS)}
Progress insights are designed to report state without surveillance. The student can set task goals and consented thresholds for releasing summaries, which enables the assistant to prepare a curated progress report when a threshold condition is met. 
The supervisor sees summaries for their own students only, while the GRS sees just the consented aggregate signals. This approach supports institutional assurance and planning while keeping control firmly with people.

\paragraph{Progress insight mechanism}
The assistant tracks task states within the student's private context store, where each task, if configured, has a clear, student-set goal and threshold. These thresholds can be based on metrics such as paper draft completeness, literature reviewed, or experimental results analysed. When a threshold is reached, the system compiles a student-curated summary complete with links to the relevant artefacts. Only the summary and its associated links are released under the chosen rule. For all other outputs, student-initiated forwarding remains the default method of sharing.

\paragraph{Stakeholder theory lens}
Legitimate claims are better served because support becomes more timely for the student, time is protected for the supervisor, and policy is applied and checked where decisions actually occur for the GRS. Furthermore, power does not shift to the tool, as veto and escalation remain entirely human functions. Urgency comes to be shaped by readiness signals rather than by the sheer volume of notifications. The assistant acts as mediator and record, not as arbiter.

\paragraph{Governance note}
All behaviour patches, logs, and insights operate under the core principles of consent, minimal data retention, and role-based access. In addition, pastoral signals are strictly limited to signposting with clear escalation rules. These combined limits ensure the tool's operation aligns with the principles of academic integrity and institutional program policy.

\subsection{Centrality of Human Supervision}

The rapid adoption of LLMs in doctoral research is undeniable. They have already been shown to be powerful tools for key PhD tasks like brainstorming, literature review, and even coding support \cite{Dai2023}. But to see this efficiency as a potential replacement for human supervision is to mistake the tool for the task. From a philosophical standpoint, a doctorate's goal is not the production of a document or published papers, but the ontological transformation of a student into a scholar. This formative journey relies on uniquely human interactions that an LLM, by its design, cannot replicate.

Expertise runs on tacit knowledge \cite{polanyi1966tacit}. It is the intuitive, embodied wisdom a scholar can enact but cannot fully state. This is the 'feel' for a flawed experiment, the instinct for a promising line of inquiry, the fine-grained grasp of a field’s politics. An LLM is trained primarily on digitised text rather than life experience. It can compress and retrieve vast corpora, yet it does not have the supervisor’s embodied history of judgement. Such knowledge cannot be fully captured in prompts, or proposed patches, and is formed through supervised practice, feedback, and apprenticeship.

Doctoral education is also an act of socialisation. It is a process of moving from the edge of a discipline to its core, an idea central to Lave and Wenger's ``communities of practice" \cite{lave1991situated}. The supervisor is the student's guide and gatekeeper in this journey. They vouchsafe the student's legitimacy, make introductions, and model the very disposition of a scholar. An LLM-driven AI assistant is not a member of this community and cannot ever be. It cannot grant a student a legitimate place at the table because it has no seat there itself. 
That students turn to AI for ``surrogate emotional support" is telling \cite{Boyd2025}. It reveals a deep need for validation and encouragement, yet an AI can only simulate empathy. It cannot build resilience through shared struggle or foster the intellectual courage required to pursue a risky idea. Meanwhile, ethical scholarship is not an algorithm. It is a disposition, learned through dialogue and modelled by a mentor.
LLMs are becoming indispensable co-pilots in the doctoral journey, which must now be used to 'keep up', handling the explicit and streamlining the mechanical \cite{Iatrellis2025}, but the human supervisor remains the irreplaceable navigator. They are the conduits of tacit knowledge, the gatekeepers to the community of practice, and the mentors of the whole person. To ignore this philosophical dimension is to reduce the doctorate to a technical exercise, stripping it of the very process that forges a scholar.

\subsection{Technosolutionism: The LLM as a Tool, Not a Panacea}

Technosolutionism holds that complex social and pedagogical problems can be solved primarily by technological means. We position this framework in direct opposition to this notion. We recognise that the challenges facing doctoral education, from student mental health and overload to supervisor capacity, are structural, not merely technical \cite{Selwyn2014, Williamson2017}. Therefore, we do not present the AI assistant as a cure for all these systemic issues, although it can alleviate many. Rather, we propose a bounded and carefully scoped tool that addresses specific pain points: feedback latency, progress management, information overload and synthesis, and the initial hurdles of academic writing. 

Our approach advocates subordinating technology to pedagogy. Implicit in our proposals is a system that embeds transparency and academic integrity, requiring that the AI assistant's contributions remain disclosed and auditable. We limit its role in affective support to offering reflective prompts, never high-stakes judgment calls or critical pastoral care, and instead, we suggest clear pathways for referral to human support. This aligns our work with the principles of critical digital pedagogy, which foregrounds human agency and treats technology as a means, not an end \cite{Stommel2020}.

Furthermore, we maintain that any implementation embodies the tenets of responsible innovation via reflexivity and inclusion, by co-designing and continuously evaluating the tool with the very community it serves \cite{Stilgoe2013}. We therefore measure success not by engagement with the tool itself, but by its impact on shifting the needle on PhD completion rates and ultimately meaningful pedagogical outcomes: the quality of scholarly work, progress toward milestones, and the enrichment of the human supervisory relationship. Thus, this framework positions the AI assistant as one component within a broader ecosystem of institutional change, offering a powerful assistant that serves, rather than supplants, the indispensable human core of doctoral mentorship.

\subsection{Ethical Implications and Risks}
\label{sec:EthicalImplications}

Our anchors are Academic Integrity principles and Stakeholder Theory. With an orchestrated AI assistant in the loop, \emph{judgement} stays with people; what changes is where information is prepared and when it arrives. Several risk classes follow from this reconfiguration.

\textbf{Relational risks: } Dialogue and tacit mentoring can thin out if roles are unclear. We therefore keep authorship and final approval with candidates and supervisors, require disclosure of assistant use on substantive work, and scope behaviour patches to a student and topic so they encode pedagogy without over-standardising practice. Pastoral screening is {signposting only} under consent and escalation rules; the assistant never delivers counselling or diagnosis. We avoid metric fixation by keeping threshold-based progress summaries student-set, revocable, and off by default, and by requiring that summaries complement rather than replace supervisor judgement.

\textbf{Informational risks: } Handling drafts, data, and correspondence through tools raises confidentiality and IP exposure. This can be mitigated with approved tooling, data minimisation, role-based access to the {context store}, limited retention, and explicit provenance: retrieved sources, tool traces, and patch history accompany outputs so supervisors can reconstruct how a recommendation was produced. Bias concerns persist; policy-relevant outputs must be grounded via RAG to authoritative corpora, with periodic checks for disparate impact. Student-set progress and completion thresholds can be gamed; we log threshold edits, show patch and tool traces, and run periodic low-support checks of competence, where the assistant is limited or off and the student explains or reproduces key steps.

\textbf{Technical risks.}
Ease invites over-reliance, and model updates, prompt injection, or tool faults can degrade performance.
We therefore suggest hardening the retrieval and tool layers against {indirect prompt injection} and {data or corpus poisoning} in RAG.
We require human-in-the-loop review for high-stakes actions and versioned policies in the GRS \emph{policy index}.
Where safety is critical we use ensembles or self-consistency runs, and we log failure modes for post-hoc review.

\textbf{Learning risk and the fluency trap.}
LLMs boost fluency and remove friction, which can create an {illusion of learning} if effortful cognitive work is displaced.
The risk profile is classic and results in reduced retrieval and self-explanation, inflated confidence, and performance that collapses when support is absent.
We design against this with interaction patterns that elicit retrieval, explanation, and decision-making rather than answers on demand.
First, supervisor patches can be designed tp steer the assistant toward a Socratic, question-led style at higher cognitive levels, so the student explains, justifies, and contrasts rather than accepts an output.
Second, the assistant can run brief, low-stakes oral checks on request or at planned points, where help is limited and the student answers aloud to explain a result, trace a proof, or walk through code.
Speaking answers recruits articulatory–motor, auditory, and somatosensory systems in addition to language areas, creating a richer multimodal trace; this well-documented {production effect} yields stronger encoding than silent study.
Third, the private loop includes retrieval practice schedules (quick self-quizzes on prior material), spaced follow-ups, and light interleaving across topics.
Lastly, prompts can be designed that require self-explanations at key steps and ask for decisions with explicit rationales, which alongside the other mitigations would support durable learning.


\section{Discussion}

The genie is out of the bottle. PhD students are already heavily using and increasingly relying on LLMs and ancillary technologies across all of their research and study tasks. Doctoral supervisors are likewise using LLMs for assisting with supervision tasks like conducting reviews of their students' work and for providing feedback. Institutional Graduate Research Schools are adopting them to streamline administration. None of this can be denied. There is no turning back, and the question is whether we are willing, risks notwithstanding, to embrace these technologies, unify them as proposed here, and steer their use with oversight. No technological breakthroughs are required to build an agentic orchestration layer. Rather, the task is to assemble and expose existing capabilities around LLMs through a unified interface.

This reframes supervision from an opaque craft to a practice that is transparent and improvable.
Behaviour patches and logs make tacit preferences explicit at the point of use.
The \textit{pedagogic relation} benefits because routine scaffolding moves to the assistant and meetings focus on judgement and identity work.
The \textit{policy–compliance relation} benefits because guidance arrives more efficiently through a GRS-managed policy index that also  supports corpus-level querying by policy authors via RAG, conflict checks, change tracing, and full logging, with searches limited to the policy corpus.
The \textit{capacity–assurance relation} benefits because supervisors receive student-curated progress summaries that are triggered by student-set goals and thresholds or issued on request, with optional, revocable auto-send, and with the GRS seeing consented aggregate signals only. Meanwhile, the supervisor remains the final arbiter and the assistant is a mediator and record, not a stakeholder.

Beyond the risks already identified, arguably, the most subtle and consequential one to highlight is the fluency trap, leading to the illusion of learning for students (and also the supervisors). 
The greatest challenge for this new, AI-enhanced supervision and the next generation of scholars in general, therefore, may not just be how to integrate a powerful tool, but to teach students the wisdom to know when to use it, when to question it, and when to turn it off and simply think.

\paragraph*{Study limitations and future work.}
This is an ex-ante, literature-grounded design rather than an empirical evaluation. Models and tools change fast and may outpace parts of the analysis. We centre three primary stakeholders and do not model external examiners, funders, or industry partners. Our proposal is generic across disciplines, and because supervision practices and AI uptake differ by field, its utility will vary across subjects.
Future work should implement the orchestration layer and run field trials across disciplines to test utility, safety, and auditability under real workflows. Longitudinal studies should track learning gains, transfer, supervision dynamics, and completion rates, with attention to online and part-time cohorts. Comparative evaluations should benchmark retrieval, routing, inference-time checks, multimodality, and audit features against human-only baselines. Governance research should assess disclosure, privacy, and review load and link these to assessment redesign and integrity policies.


\section{Conclusion}
\label{sec:Conclusion}

This study presented a design-science blueprint for integrating an orchestrated AI assistant and co-pilot into doctoral supervision as a socio-technical mediator rather than a replacement supervisor. The artefact is grounded in Stakeholder Theory and bounded by Academic Integrity, and it maps recurrent supervision pain points to capabilities with a transparent severity–mitigability triage. The workflow composes available parts into one accountable system that includes retrieval-augmented generation, mixture-of-experts routing, ensembles with self-consistency, multimodality, a GRS-managed policy index, a student context store, and behaviour patches that make tacit guidance explicit while keeping authorship and final judgement with people.
Our analysis supports a simple sequencing principle. Use the assistant to make the smooth parts of supervision smoother and to automate routine tasks that consume scarce human capacity. Redirect the recovered time and attention to high-severity relational challenges that only humans can address. Naturally, risks remain. The key risk may ultimately be the illusion of learning when fluency displaces effortful cognitive processing in learners. This is a design risk rather than an inevitability, and we outline mitigations alongside other risks.

Claims are \textit{ex ante} and conceptual and no new empirical data were collected. Future work should implement the orchestration layer and run field trials across disciplines to measure utility, safety, auditability, learning gains, supervision dynamics, and completion outcomes, with attention to online and part-time cohorts. Institutions can act now by adopting the governance stance and design rules set out here so existing capabilities are unified under oversight and human authority is preserved while day-to-day friction is reduced.

\bibliographystyle{unsrtnat}

\appendix

\clearpage

\section{Supplementary Material}


\section*{Methods: Severity scoring — workings, sources, and assumptions}

This section expands and presents the calculations, assumptions, justifications, and all the workings behind the derivation of the severity score in Table~2 in the manuscript.
From Table~2, we compute a per-issue \emph{severity} score \(S\in\{1,2,3\}\) as
\[
S=\mathrm{round}\!\Big(\frac{P+C}{2}\Big),
\]
where \(P\) is a banded estimate of \textbf{prevalence} and \(C\) is a banded estimate of \textbf{consequence}. We set \(P=1\) for Low prevalence (\(<20\%\)), \(P=2\) for Moderate prevalence (\(20\text{--}40\%\)), and \(P=3\) for High prevalence (\(>40\%\)) based on recent large surveys or reviews when available. We set \(C=1\) for Low consequence (nuisance/admin friction), \(C=2\) for Moderate consequence (impairs progress/quality), and \(C=3\) for High consequence (linked to intent-to-leave/time-to-degree or serious wellbeing/ethical risk; typically supported by odds ratios \(\geq 2\) or convergent evidence). Where required, we state assumptions and confidence. Citations below use the same keys as in the manuscript.

\vspace{0.5em}
\noindent\textbf{Key sources used across issues.} Global PhD indicators on finances, publish pressure, harassment, and AI guidance needs \cite{nordling2025money}. Mental-health impact on attrition risk \cite{busch2024behind}. Discipline-agnostic supervision feedback synthesis \cite{bearman2024enhancing}. Challenge–hindrance framing \cite{acharya2024challenge}. Power/working-conditions study \cite{moshtari2025supervisors}. Relationship style findings \cite{mavrogalou2024relationship}. Information-overload reviews \cite{arnold2023dealing,shahrzadi2024causes}. Writing support for EAL \cite{khalifa2024using}. Where a source reports category presence rather than exact percentages, we conservatively assign Moderate prevalence.

\subsection*{Per-issue calculations (matching Manuscript Table~2)}

\paragraph{Mental-health strain linked to research/teaching pressures and progress reporting \cite{busch2024behind,levecque2017work}.}
\textbf{Prevalence:} \(P=2\) (Moderate). Cross-country estimates place clinical-range distress or severe anxiety/depression for sizeable minorities of PhD students \cite{levecque2017work}.  
\textbf{Consequence:} \(C=3\) (High). Severe anxiety/depression associated with \(\sim\)3–5\(\times\) higher odds of considering leaving \cite{busch2024behind}.  
\textbf{Computation:} \(S=\mathrm{round}((2+3)/2)=\mathrm{round}(2.5)=3\).  
\textbf{Assumptions/confidence:} Confidence High; prevalence banded conservatively at Moderate.

\paragraph{Harsh criticism and unreasonable expectations \cite{busch2024behind}.}
\textbf{Prevalence:} \(P=2\) (Moderate). Reported as common detrimental experiences, but not quantified as a majority in cited work.  
\textbf{Consequence:} \(C=3\) (High). Identified as among the most detrimental research aspects for anxiety/depression \cite{busch2024behind}.  
\textbf{Computation:} \(S=\mathrm{round}((2+3)/2)=3\).  
\textbf{Assumptions/confidence:} Confidence Medium–High; consequence banded High per qualitative weight.

\paragraph{Pressure to publish and “publish-or-perish” expectations \cite{nordling2025money}.}
\textbf{Prevalence:} \(P=3\) (High). “Need to get published” reported by \(\sim\)41\% as a major concern.  
\textbf{Consequence:} \(C=3\) (High). Consistent linkage to stress, satisfaction, and persistence pressure; influences continuation decisions indirectly.  
\textbf{Computation:} \(S=\mathrm{round}((3+3)/2)=3\).  
\textbf{Assumptions/confidence:} Confidence High.

\paragraph{Teaching burden with limited training \cite{busch2024behind}.}
\textbf{Prevalence:} \(P=2\) (Moderate). Teaching responsibilities common, but \emph{burden + low training} as a severe problem is not majority-quantified.  
\textbf{Consequence:} \(C=2\) (Moderate). Reported to exacerbate anxiety, but with heterogeneous exposure/mitigation.  
\textbf{Computation:} \(S=\mathrm{round}((2+2)/2)=2\).  
\textbf{Assumptions/confidence:} Confidence Medium; we avoid double-counting mental-health effects already captured above.

\paragraph{Emotional load, boundary strain, and reputational risk (Supervisor) \cite{li2025phd,bearman2024enhancing}.}
\textbf{Prevalence:} \(P=2\) (Moderate). Recurring theme in qualitative studies; limited large-\(N\) quantification for supervisors.  
\textbf{Consequence:} \(C=2\) (Moderate). Impacts supervisor wellbeing and professional standing; indirect effects on candidate experience.  
\textbf{Computation:} \(S=\mathrm{round}((2+2)/2)=2\).  
\textbf{Assumptions/confidence:} Confidence Medium; system-level knock-on effects plausible but under-measured.

\paragraph{Feedback timeliness and feedforward quality \cite{bearman2024enhancing}.}
\textbf{Prevalence:} \(P=2\) (Moderate). Identified widely as a supervision problem; not uniformly present across all dyads.  
\textbf{Consequence:} \(C=3\) (High). Absence/poor quality linked to distress and formal failure at milestones in case evidence \cite{bearman2024enhancing}.  
\textbf{Computation:} \(S=\mathrm{round}((2+3)/2)=3\).  
\textbf{Assumptions/confidence:} Confidence Medium–High; we treat milestone failure as an attrition-risk signal.

\paragraph{Power asymmetry in feedback and working conditions \cite{bearman2024enhancing,moshtari2025supervisors}.}
\textbf{Prevalence:} \(P=3\) (High). Structural feature of supervision; frequently reported across contexts.  
\textbf{Consequence:} \(C=3\) (High). Associated with hostile climates and wellbeing harms; governance salience is high.  
\textbf{Computation:} \(S=\mathrm{round}((3+3)/2)=3\).  
\textbf{Assumptions/confidence:} Confidence High.

\paragraph{Challenge–hindrance demands and bureaucratic load \cite{acharya2024challenge}.}
\textbf{Prevalence:} \(P=2\) (Moderate). Hindrance demands recur across programmes.  
\textbf{Consequence:} \(C=2\) (Moderate). Impede progress and wellbeing; typically not sole drivers of attrition.  
\textbf{Computation:} \(S=\mathrm{round}((2+2)/2)=2\).  
\textbf{Assumptions/confidence:} Confidence Medium.

\paragraph{Supervisory style inconsistency and relationship strain \cite{mavrogalou2024relationship}.}
\textbf{Prevalence:} \(P=2\) (Moderate). Relationship strains are common but not universal.  
\textbf{Consequence:} \(C=2\) (Moderate). Although uncertain style is a strong predictor of poor mental-health outcomes in one model, generalisable causal evidence remains limited.  
\textbf{Computation:} \(S=\mathrm{round}((2+2)/2)=2\).  
\textbf{Assumptions/confidence:} Confidence Medium; we avoid inflating \(C\) pending broader replication.

\paragraph{Synthesis and academic writing load (esp. EAL) \cite{khalifa2024using}.}
\textbf{Prevalence:} \(P=2\) (Moderate). Writing load is universal, but \emph{problematic} load concentrates among subgroups (e.g., EAL, novices).  
\textbf{Consequence:} \(C=2\) (Moderate). Affects pace/quality more than direct attrition.  
\textbf{Computation:} \(S=\mathrm{round}((2+2)/2)=2\).  
\textbf{Assumptions/confidence:} Confidence Medium–High for \(P\) banding; outcome effects vary by support.

\paragraph{Funding precarity and financial stress \cite{nordling2025money}.}
\textbf{Prevalence:} \(P=3\) (High). Financial pressure is the top concern for \(\sim\)42\%.  
\textbf{Consequence:} \(C=3\) (High). A majority indicate inflation affects continuation decisions (proxy for attrition risk).  
\textbf{Computation:} \(S=\mathrm{round}((3+3)/2)=3\).  
\textbf{Assumptions/confidence:} Confidence High.

\paragraph{Harassment and hostile climates \cite{nordling2025money}.}
\textbf{Prevalence:} \(P=3\) (High). \(\sim\)43\% report discrimination/harassment.  
\textbf{Consequence:} \(C=3\) (High). Direct threat to wellbeing and belonging; strong persistence risk.  
\textbf{Computation:} \(S=\mathrm{round}((3+3)/2)=3\).  
\textbf{Assumptions/confidence:} Confidence High.

\paragraph{Information overload in the literature \cite{arnold2023dealing,shahrzadi2024causes}.}
\textbf{Prevalence:} \(P=2\) (Moderate). Reported by substantial minorities in scholarly settings.  
\textbf{Consequence:} \(C=2\) (Moderate). Related to strain/burnout in some contexts, but direct doctoral attrition evidence is limited; we conservatively set Moderate.  
\textbf{Computation:} \(S=\mathrm{round}((2+2)/2)=2\).  
\textbf{Assumptions/confidence:} Confidence Medium; we avoid double-counting mental-health S=3 category.

\paragraph{Ambiguity about acceptable AI use}
\textbf{P=2 (Moderate):} Although \(\sim\)64\% of students report wanting more guidance on AI \cite{nordling2025money}, we scope prevalence to cases where ambiguity materially impairs decisions, creates friction in supervision, or credibly raises integrity risk; hence Moderate rather than High. 
\textbf{C=2 (Moderate):} Ambiguity generates anxiety about inadvertent misconduct, uneven expectations across labs, and mistrust \cite{jin2025generative}, but direct links to attrition are not yet strongly quantified.
\textbf{Calculation:} \(S=\text{round}((2+2)/2)=2\).
\textbf{Assumptions/Confidence:} We avoid equating ``desire for guidance” with “problematic ambiguity” to prevent prevalence inflation; consequence focuses on integrity and relational friction rather than attrition. \emph{Confidence: Medium} (solid prevalence signal, developing consequence evidence).

\paragraph{Supervision capability \& milestone quality support}
\textbf{P=1 (Low):} As a \emph{recognised} pain point, programme-level enablement sits behind dyadic issues such as finance or feedback in frequency. 
\textbf{C=1 (Low):} GRS-level levers (rubric packs, exemplar banks, CPD, panel calibration) improve consistency but rarely drive attrition on their own \cite{bearman2024enhancing}.
\textbf{Calculation:} \(S=\text{round}((1+1)/2)=1\).
\textbf{Assumptions/Confidence:} Scoped to research-led doctorates where GRS cannot access private supervisory comments; effects are indirect and contingent on local uptake. \emph{Confidence: Medium} (clear role boundaries, limited causal evidence for direct programme outcomes).

\subsection*{Methods: Risk Assessment Rationale}

Supplementary Table\ref{tab:risk-justifications} expands on the rationale used to derive the risk column in Table~2 found in the main manuscript.
Low means limited new exposure with strong guardrails.
Medium means meaningful exposure that is mitigable in routine use.
High means credible exposure in sensitive domains even when guardrails are present.

\begin{table}[hbt!]
\centering
\caption{Risk justifications for issues in main Manuscript Table~2}
\label{tab:risk-justifications}
\begin{tabularx}{\textwidth}{>{\raggedright\arraybackslash}p{0.3\textwidth} >{\centering\arraybackslash}p{0.08\textwidth} >{\raggedright\arraybackslash}X}
\toprule
\textbf{Issue} & \textbf{Risk} & \textbf{Rationale} \\
\midrule
Mental-health strain linked to research or teaching pressures & H &
False reassurance, misclassification, and implicit surveillance remain plausible failure modes even with signposting-only controls; human escalation reduces harm but cannot remove it in sensitive contexts. \\
Harsh criticism and unreasonable expectations & M &
Tone scaffolds reduce the chance of abrasive feedback; residual risks of mis-toning and homogenised voice are manageable with supervisor veto and patch updates. \\
Pressure to publish and quantity expectations & H &
Acceleration and polish can amplify quantity pressure and raise integrity risk in competitive settings; disclosure prompts help, yet incentives and power dynamics keep exposure high. \\
Teaching burden with limited training & L &
Assistant-authored teaching aids run under approved templates and local storage, and the feature set does not introduce material privacy or agency risks. \\
Emotional load and boundary strain for supervisors & M &
Boundary scripts and agenda builders reduce friction, yet escalation choices still carry exposure because they involve context-sensitive human judgement. \\
Feedback timeliness and feedforward quality & M &
Automated comments and next-step plans reduce delay; main risks are over-reliance and occasional mis-advice, which logs, grounding, and behaviour patches mitigate. \\
Power asymmetry in feedback and working conditions & H &
Logs and insight flows can entrench power if misused; guardrails limit scope and access, yet the structural risk remains high. \\
Challenge–hindrance demands and bureaucratic load & M &
Automation lowers effort on forms and deadlines; process rigidity and rule drift remain possible and require periodic review. \\
Supervisory style inconsistency and relationship strain & L &
Compacts, minutes, and action tracking add clarity and predictability while introducing little new exposure. \\
Synthesis and academic writing load & M &
Grounded drafting and checking reduce error and effort; plagiarism and voice drift remain credible risks, contained by provenance, disclosure, and assessment design. \\
Funding precarity and financial stress & L &
Search and drafting tools improve applications but do not change availability of funds; new exposure is limited under standard controls. \\
Harassment and hostile climates & H &
False confidence in automated triage and chilling effects are possible in sensitive reporting; human enforcement and clear policy are required and exposure remains high. \\
Information overload in the literature & M &
Curated RAG with provenance and contrastive summaries reduces overload and hallucination risk; residual risks are anchoring on early sources and missed breadth, which audit trails and periodic breadth checks can manage. \\
Ambiguity about acceptable AI use and integrity & M &
Policy-aware prompts and exemplars improve compliance at the point of writing; drift and uneven uptake persist across programmes, so exposure remains moderate. \\
Supervision capability and milestone quality support & H &
Student-set goals and consented thresholds raise mitigability and improve meeting focus; gaming and metric fixation are credible if thresholds are mis-set or enforced without judgement, so exposure stays high and requires audit trails and periodic low-support checks. \\
\bottomrule
\end{tabularx}
\end{table}

\end{document}